\documentclass[aps,reprint,twocolumn,showkeys,showpacs,superscriptaddress]{revtex4-1}

\usepackage{amsmath,amssymb}
\usepackage{graphicx}% Include figure files
\usepackage{dcolumn}% Align table columns on decimal point
\usepackage{bm}% bold math
\usepackage{color}
\usepackage{soul}
\usepackage[normalem]{ulem}
\usepackage{hyperref}

\usepackage{natbib}
\setcitestyle{super}
\begin{document}
\title{Nonequilibrium Excitations and Transport of Dirac
Electrons in
Electric-Field-Driven Graphene}
\author{Jiajun Li}
\affiliation{Department of Physics, State University of New York at
Buffalo, Buffalo, New York 14260, USA}
\affiliation{Department of Physics, Friedrich-Alexander-Universit\"{a}t Erlangen-N\"{u}rnberg, Erlangen 91054, Germany}
\author{Jong E. Han}
\affiliation{Department of Physics, State University of New York at
Buffalo, Buffalo, New York 14260, USA}
\date{\today}

\begin{abstract}

We investigate nonequilibrium excitations and charge transport in
charge-neutral graphene driven with dc electric field by using the
nonequilibrium Green's function technique. Due to the vanishing Fermi
surface, electrons are subject to non-trivial nonequilibrium excitations
such as highly anisotropic momentum distribution of electron-hole pairs,
an analog of the Schwinger effect. We show that the electron-hole
excitations, initiated by the Landau-Zener tunneling with a superlinear
$IV$ relation $I \propto E^{3/2}$, reaches a steady state dominated by the
dissipation due to optical phonons, resulting in a marginally sublinear
$IV$ with $I \propto E$, in agreement with recent experiments. The linear
$IV$ starts to show the sign of current saturation as the graphene is
doped away from the Dirac point, and recovers the
semi-classical relation for the saturated velocity. We give
a detailed discussion on the nonequilibrium charge creation and the
relation between the electron-phonon scattering rate and the electric
field in the steady-state limit. We explain how the apparent Ohmic $IV$ is
recovered near the Dirac point. We propose a mechanism where the
peculiar nonequilibrium electron-hole creation can be utilized in an infra-red device.

\end{abstract}

\pacs{72.80.Vp,73.50.Fq}
\maketitle

\section{Introduction}

As a prototypical monolayer material, graphene has attracted much
attention in the past decade for its extraordinary mechanical and
electronic properties~\cite{Novoselov666,neto09,dassarma11}. The
extraordinarily high mobility up to $10^5$ cm\textsuperscript{$2$}/V s and the
current density up to $10^9$A cm\textsuperscript{$-2$} in graphene\cite{moser07}
make the system a prominent building element in nanoelectronics.
The peculiar linear dispersion relation of the
band structure at the charge-neutrality point has fascinated the physics
community ever since graphene could be mass produced, due to its
novel relativistic analog in a solid state system. The vanishing gap out
of the honeycomb lattice has the most attractive aspects in both worlds
of semiconductor and metal: the maneuverability of the semiconductor and
the cleanness of metal.

Among many fundamental questions unique to graphene, we explore here the
role of the Dirac point in electronic transport.  The dc transport
experiments have shown the electric current in graphene tends to
saturate under a high electric field of several tens of
kV/cm\cite{meric08,barreiro09,dorgan10,ramamoorthy15}. Theoretical and
experimental studies have indicated that the tendency is due to the
interaction between electrons and optical
phonons\cite{meric08,barreiro09,chauhan09,dasilva10,perebeinos10,fang11}.
Despite the progress, most theoretical studies have been limited to
the Boltzmann transport theory and applied mostly to samples with high
electron densities. We present a detailed study of electronic
excitations and the transport close to the Dirac point by using the
microscopic calculation based on the Keldysh Green's function method. We
employ the formulation recently developed for the dissipative
steady-state nonequilibrium under a dc electric field~\cite{jiajun-prl},
applicable for experimental dc transport measurements.

Recently, various transport mechanisms in graphene have been proposed.
Boltzmann transport theory~\cite{meric08} and a streaming
model~\cite{fang11} have been successful to describe the behavior of the
velocity saturation in the limit of large current density. However,
in the charge-neutrality limit, the peculiarity of the Dirac point demands
proper quantum mechanical treatment of the nonequilibrium effects.
It has been argued that the Landau-Zener tunneling~\cite{zener} should
play an important role in the nonequilibrium state near the Dirac 
point~\cite{danneau08,Kao10,Miao1530,oladyshkin17,Rosenstein10,Vandecasteele10,kane2015,higuchi17}, 
where the electric field creates excitations in pairs of electron and
hole, dramatically changing its equilibrium electronic properties. This
effect can be considered a solid-state analog of the Schwinger
effect~\cite{schwinger51,dora10,allor08,gourdeau15}. However, the main
consequence from the
theory on the current-voltage relation, $I\sim V^\alpha$ with the
exponent $\alpha$ greater than 1, has been inconsistent with majority of the
measured dc transport in graphene\cite{barreiro09,ramamoorthy15,Yang18} where the
current saturation gradually crosses over to a linear or marginally
sublinear $I$-$V$ relation ($\alpha\approx 1$).

To investigate the transport mechanism, we start from a conventional
quantum mechanical tight-binding model for graphene lattice with
coupling to on-site phonons to simulate the optical phonons. In
addition, we implement the dissipation mechanism in the form of fermion
baths~\cite{jong-prb,jiajun-prb} to mimic the dissipation into an
infinite medium which is essential to establish a rigorous steady-state
nonequilibrium limit within the defined Hamiltonian. We use the Keldysh
formalism~\cite{jiajun-prl} to obtain the steady-state Green's function
(GF) and then transport quantities. The calculation confirms the
semi-classical behaviors away from the Dirac point, predicted by the
Boltzmann transport theory. In the charge-neutrality limit, the
excitation population is linearly proportional to the electric field in
the presence of the coupling to optical phonons, while the drift
velocity saturates to about 50\% of the Fermi
velocity~\cite{ramamoorthy15}.  Despite the reversed role of the charge
excitation and the drift velocity in the Drude metal, an apparent Ohmic
relation $I\propto V$ is established. Furthermore, the electron-hole
pair creation is strongly anisotropic occupying the same region of the
momentum space in the upper and lower Dirac cones, which makes the
system a strong candidate for an infra-red switching device.

The paper is organized as follows. In Sect. II, the model and the
methodology are detailed. Especially, the construction of the lattice
summation in the nonequilibrium Dyson equation is explained. In Sect.
III, we first discuss the Landau-Zener effect in the absence of the
optical phonons. We then discuss the effect of the phonons on
$IV$-relation and the excitation distribution. We give detailed analysis
and microscopic justification for the saturated transport limit through
electron-phonon (el-ph) coupling. In Sect. IV, we summarize and
speculate possible device application by exploiting the peculiar
excitation spectrum in graphene.

\section{Formulation}

\subsection{Model}

We introduce a dissipative lattice model which consists of tight-binding
Hamiltonian coupled to bath systems with open boundary. We solve the
problem strictly within the given Hamiltonian according to the Keldysh
formalism, and as established
previously~\cite{jong-prb,jiajun-prb,jiajun-prl,jiajun-nano}, the
coupling to infinite degrees of freedom facilitates the infinite-time
limit for steady-states under a dc electric field. We use fermion baths
to mimic the continuous medium for Ohmic dissipation. We then
include the inelastic scattering mechanism provided by optical phonons
which will be shown to be crucial to understand the transport phenomena
in graphene. Scattering due to Coulomb interaction is not considered
in this work. While the Coulomb interaction effectively assists the relaxation of electronic energy
and momentum on the femtosecond time scale, unlike electron-phonon coupling, its effect in graphene at the limit of strong dc electric field has not been established experimentally. We will come back to this
point in Sect. III C.

\begin{figure}
\centering
\includegraphics[scale=0.6]{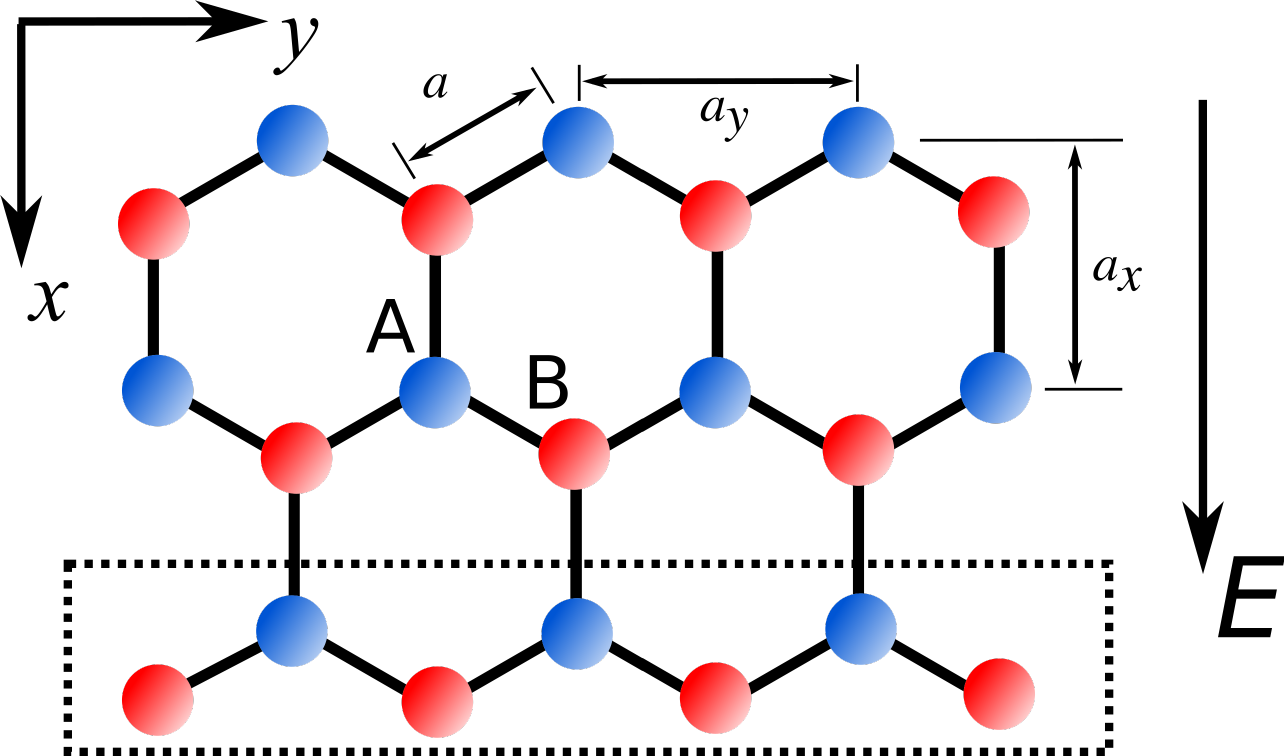}
\caption[Tight-binding model of graphene]{Tight-binding model of
graphene. The hexagonal lattice structure is composed of two triangular sublattices
labeled as $A$ (blue) and $B$ (red).
 The dashed line encircles a ``zig-zag" row perpendicular to the field direction.}
\label{grphmodel}
\end{figure}

The Hamiltonian is broken up as follows.
\begin{align}
H=H_\text{TB}+H_\text{bath}+H_\text{ph}+H_\text{E}.
\end{align}
$H_\text{TB}$ is the tight-binding model of graphene defined on a
honeycomb lattice as shown in Fig.~\ref{grphmodel}. $H_\text{bath}$ is
the coupling to fermion baths, and $H_\text{ph}$ is the coupling to optical
phonons. $H_\text{E}$ is the energy shift of the tight-binding and bath
orbitals due to the external electric field $E$.

The tight-binding Hamiltonian $H_\text{TB}$ is defined on a monolayer honeycomb
lattice which is a system of interlaced triangular sublattices $A$ and
$B$ as shown in Fig.~\ref{grphmodel}. The Hamiltonian is written as
\begin{align}
H_\text{TB}=-\gamma\sum_{\langle\bm{r}\bm{r}'\rangle}d^\dag_{\bm{r}'}d_{\bm{r}},
\end{align}
with the tight-binding parameter $\gamma$ and the electron operator
$d_{\bm{r}}$ defined on lattice vectors $\bm{r}$. The lattice summation
is restricted to nearest neighbors, i.e., $\bm{r}$ and $\bm{r}'$
should be on different sublattices. The tight-binding
parameter $\gamma$ is typically $3$ eV for graphene~\cite{neto09}. 
We ignore the spin degree of freedom.

The coupling to the fermion reservoirs in the Hamiltonian, $H_\text{bath}$, is of the following form
\begin{align}
H_\text{bath}=&-\frac{g}{\sqrt{L}}\sum_{\bm{r}\alpha}\left(d^\dag_{\bm{r}}c_{\bm{r}\alpha} +\text{H.c.}\right)+\sum_{\bm{r}\alpha}\epsilon_\alpha c^\dag_{\bm{r}\alpha}c_{\bm{r}\alpha}.
\end{align}
Here, we attach a fermion continuum to each fermion site $\bm{r}$.
$c_{\bm{r}\alpha}$ is the electron operator in the continuum at the $\bm{r}$ site with
the continuum index $\alpha$ at energy $\epsilon_\alpha$. For
simplicity, we assume that the energy spectrum of $\epsilon_\alpha$ has
a constant density of states. Then the hybridization of the
$\bm{r}$ site to the reservoir is given by the damping parameter
$\Gamma$ as $\Gamma=L^{-1}\pi g^2\sum_{\alpha}\delta(\epsilon_\alpha)$
with the volume normalization by $L$ [$\delta(\epsilon_\alpha)$ is a Dirac $\delta$ function]. This provides a physical mechanism
to dissipate excess energy created by external fields and enables a
steady state with a finite current. The scattering time due to acoustic
phonons scales linearly\cite{hwang08,pietronero80} with electron energy
and is up to about 5 ps\textsuperscript{$-1$} within the Dirac
cone\cite{fang11}, which corresponds to $\Gamma/\gamma\sim
5\times10^{-4}$ in our model. We emphasize that the dissipation of the
model is not inside the leads, as often considered in nano-junction
models, but comes from the \textit{bulk} dissipation where the energy
relaxation occurs over the whole system. 

Electrons in the monolayer lattice interact with both acoustic and
optical phonons. The former has zero energy gap and can be excited by
arbitrarily small energy. In the low-field regime, interaction with
acoustic phonons provides a fundamental channel of electronic scattering
and energy dissipation. On the other hand, the optical phonons have a
large gap $\hbar\omega_\text{ph}$ and interact strongly with electrons
only through higher-energy processes. It is argued that the dissipation
by acoustic phonons is described by considering the exactly
soluble fermion-reservoir model~\cite{jong-prb}. In the regime of small
dissipation, this model reproduces successfully the Boltzmann transport
theory and gives the correct linear response
behavior~\cite{jiajun-prb,jiajun-prl}. This motivates us to consider the
graphene lattice coupled to fermion reservoirs at each lattice site. 

In the strong-field transport, inclusion of inelastic scattering becomes
crucial and we consider the optical phonons at
frequency $\omega_\text{ph}\approx 150$ meV
($\hbar=1$ unit is used)\cite{perebeinos10}. Each lattice site couples to an
independent (optical) phonon bath with coupling constant $g_{\rm ep}$. The
el-ph coupling takes the typical Holstein model as
\begin{align}
H_\text{ph}=g_{\rm ep}\sum_{\bm{r}}(a_{\bm{r}}+a^\dag_{\bm{r}})d^\dag_{\bm{r}}d_{\bm{r}}+\sum_{\bm{r}}\omega_\text{ph}a^\dag_{\bm{r}}a_{\bm{r}},
\end{align}
with $a_{\bm{r}}$ being the annihilation operator of optical phonons at position $\bm{r}$.

The last term in the Hamiltonian is the potential shift by the dc
electric field. We choose the Coulomb gauge with the static potential
$-\bm{r}\cdot\bm{E}$ ($e=1$ unit is used).
\begin{align}
H_E=-\sum_{\bm{r}}\bm{r}\cdot\bm{E}\left(d^\dag_{\bm{r}}d_{\bm{r}}+\sum_\alpha c^\dag_{\bm{r}\alpha}c_{\bm{r}\alpha}\right).
\label{hbias}
\end{align}
The summation is over the sites $\bm{r}$ with infinite size lattice. We
set the bath chemical potentials to match the level shift by the electric
field, i.e., $\mu_{\rm bath}(\bm{r})=-\bm{r}\cdot\bm{E}$, so that each
site is equivalent except for the relative energy shift. In the following
discussions, unless specifically mentioned, we reserve the chemical potential 
notation $\mu$ to that in the zero field limit.

\subsection{Recursion relations}

To solve the model, we utilize the dynamical mean-field theory
(DMFT)~\cite{RMP-NEQDMFT} within the Keldysh GF formalism,
and ignore the intersite self-energy by the el-ph
coupling\cite{georges96}. In addition to that we keep the nonequilibrium GFs, the
main difference from the equilibrium DMFT is that the lattice summation
in the DMFT cannot be performed by a sum over wavevectors
due to the bias potential. Although the lattice summation is
possible by a brute-force matrix inversion in real-space, a much more
efficient algorithm can be used in the spatially uniform limit,
as proposed in Ref.~\citenum{jiajun-prl}.

We note that the wavevector component $p_y$ perpendicular to the field
$E$ is a good quantum number, and we organize the Hamiltonian as
depicted in Fig.~\ref{grphmodel} into zig-zag rows (dashed rectangle, indexed by
$\ell$). We consider the hopping within the row parametrized by $p_y$,
and then exactly treat the inter-row
hopping through the self-energy addition ${\bf F}_\pm(\omega)$ from the
(semi-infinite) upper rows and the lower rows, respectively.

To proceed, we diagonalize the Hamiltonian within the $\ell$-th row with
the Fourier transform $d^\dag_{\ell
s}(p_y)=\frac{1}{\sqrt{N_y}}\sum_{\bm{r}\in\{\ell s\}}e^{i p_y
\bm{r}\cdot\bm{\hat{e}}_y}d^\dag_{\bm{r}}$, where the summation is
limited to all atoms inside the $\ell$-th row and with the sublattice index $s=\pm$ 
(or $A$/$B$).
$N_y$ is the number of atoms in a row for normalization. We then obtain the
transverse Hamiltonian per momentum $p_y$:
\begin{align}
H_{\text{TB}, \perp}(p_y)=-2\gamma\sum_{\ell}\cos(p_ya_y/2)d^\dag_{\ell
A}(p_y)d_{\ell B}(p_y)+\text{H.c.}
\label{perph}
\end{align}
The constant $a_y=\sqrt{3}a$ is shown in Fig.~\ref{grphmodel}. The
longitudinal Hamiltonian is
\begin{align}
H_{\text{TB}, \parallel}(p_y)=-\gamma\sum_{\ell}d^\dag_{\ell+1,A}(p_y)d_{\ell B}(p_y)+\text{H.c.}
\label{parah}
\end{align}
Now the problem is reduced to solving an effectively one-dimensional
problem parametrized by $p_y$, where a central row ($\ell=0$) is
connected to two semi-infinite chains with upper ($\ell>0$) and lower
($\ell<0$) rows via $H_{\text{TB}, \parallel}(p_y)$.

Given $p_y$, the GFs ${\bf G}^{\lessgtr}(\omega;p_y)$ at $\ell=0$ are expressed in
the $2\times 2$ sublattice space. By denoting the GF on the
edge row of the semi-infinite chains as ${\bf F}_\pm(\omega;p_y)$, we
construct the (retarded) GF as\cite{jiajun-prl,RMP-NEQDMFT}
\begin{align}
\mathbf{G}^r(\omega;p_y)^{-1}&=\omega-{\bf h}_\perp(p_y)-\mathbf{\Sigma}^r(\omega)\nonumber\\
&-\sum_{\alpha=\pm}\bm{v}_\alpha^\dag{\mathbf{F}}^{r}_\alpha(\omega+\alpha Ea_x;p_y)\bm{v}_\alpha,
\end{align}
with a given local self-energy $\mathbf{\Sigma}^r(\omega)$ by using Dyson's equation\cite{jiajun-prl}. The
intra-row Hamiltonian matrix from Eqs.~(\ref{perph}) and (\ref{hbias}) is
\begin{align}
&{\bf h}_\perp(p_y)=\begin{pmatrix}0&-2\gamma\cos(p_ya_y/2)\\-2\gamma\cos(p_ya_y/2)&-Ea/2\end{pmatrix}.
\end{align}
The matrix $\bm{v}_\pm$ is to connect the sublattices via hopping
\begin{align}
\bm{v}_+=\bm{v}^\dag_-=\begin{pmatrix}0&&\gamma\\0&&0\end{pmatrix}.
\end{align}
The GF on the semi-infinite chains $\mathbf{F}^{\lessgtr}_\alpha(\omega;p_y)$ is calculated
recursively~\cite{jiajun-prl}. By exploiting the self-similarity
between the edge and the next-to-edge rows, we have the relation
\begin{align}
\mathbf{F}^{r}_\alpha(\omega;p_y)^{-1}&=\omega-{\bf
h}_\perp(p_y)-\mathbf{\Sigma}^r(\omega)\nonumber\\
&-\bm{v}_\alpha^\dag\mathbf{F}^{r}_\alpha(\omega+\alpha Ea_x;p_y)\bm{v}_\alpha.
\label{recursion_relation}
\end{align}
Similarly, the lesser GFs can be computed with the Dyson's equations\cite{jiajun-prl},
\begin{align}
\mathbf{G}^<(\omega;p_y)&=\mathbf{G}^r(\omega;p_y)\bigg[ \bm{\Sigma}^<(\omega)\nonumber\\
&+\sum_{\alpha=\pm}\bm{v}_\alpha^\dag\mathbf{F}^{<}_\alpha(\omega+\alpha
Ea_x;p_y)\bm{v}_\alpha\bigg]\mathbf{G}^a(\omega;p_y)\nonumber \\
\mathbf{F}_\alpha^<(\omega;p_y)&=\mathbf{F}_\alpha^r(\omega;p_y)[ \bm{\Sigma}^<(\omega)\nonumber\\
&+\bm{v}_\alpha^\dag\mathbf{F}_\alpha^{<}(\omega+\alpha Ea_x;p_y)\bm{v}_\alpha]\mathbf{F}_\alpha^a(\omega;p_y).
\end{align}
We complete the DMFT loop in the usual manner. The local GF is defined as $\mathbf{G}_\text{loc}(\omega)=\frac{1}{N_y}\sum_{p_y}\mathbf{G}(\omega;p_y)$.
Once we have $\mathbf{G}_\text{loc}(\omega)$ and
$\mathbf{\Sigma}(\omega)$, we construct the (non-interacting)
Weiss-field GF $\mathbf{\mathcal{G}}(\omega)$ as
\begin{align}
\mathbf{\mathcal{G}}^{r}(\omega)^{-1}=
\mathbf{G}^{r}_\text{loc}(\omega)^{-1}+\mathbf{\Sigma}^{r}(\omega),
\end{align}
for the retarded functions. Then the Weiss-field
$\mathbf{\mathcal{G}}(\omega)$ is used to update the self-energy
$\mathbf{\Sigma}(\omega)$ which is then used to update the GFs again, as
described so far. The procedure is repeated until a convergence is
reached with the 1\% variation of current between iterations.

The self-energy has the contribution from the fermion baths and the
el-ph coupling:
$\mathbf{\Sigma}(\omega)=\mathbf{\Sigma}_\Gamma(\omega)+\mathbf{\Sigma}_\text{ph}(\omega)$.
Within the DMFT, the self-energies are diagonal in the sublattice space.
The self-energy by the fermion baths (at $\ell=0$) is given as
\begin{equation}
\mathbf{\Sigma}^r_\Gamma(\omega)=-i\Gamma,
\mathbf{\Sigma}^<_\Gamma(\omega)=-2i\Gamma\left(\begin{array}{cc}
f_0(\omega) & 0 \\
0 & f_0(\omega+Ea/2) \end{array}
\right),
\end{equation}
with the Fermi-Dirac function at the bath temperature.
The bath temperature has been chosen as $T_\text{bath}=0.01\gamma\approx 350$ K unless stated otherwise. For the el-ph
coupling, we use the approximation that the phonon GFs are not dressed
with the self-energy
\begin{eqnarray}
\Sigma^<_{\text{ph},ss}(\omega) & = & g_{\rm ep}^2\left[\mathcal{G}^>_{ss}(\omega-\omega_\text{ph})(N_\text{ph}+1)
\right. \nonumber\\
&+ & \left.\mathcal{G}^<_{ss}(\omega+\omega_\text{ph})N_\text{ph}\right],
\end{eqnarray}
with $N_\text{ph}=1/[\exp(\omega_\text{ph}/T_{\rm ph})-1]$ the
Bose-Einstein distribution and the phonon temperature $T_\text{ph}$. We assume that $\omega_\text{ph}\gg T_\text{ph}$ with
$\omega_\text{ph}\sim 1500$ K, and that the renormalization of the
phonon is not strong. 

This approximation is in line with the semiclassical pictures adopted
in previous works~\cite{meric08,barreiro09}. For very large electric
fields, the physics may be affected by the hot (optical) phonons excited
during the transport process. However, this effect is minimized if
phonons are strongly coupled with the environmental bath, and lose
energy immediately. This assumption is proven relevant 
when the graphene sample is in contact with a substrate which dissipates 
energy very efficiently, such as in the case of
hexa-boron-nitride\cite{Yang18}. In Sect. III.~E, we estimate 
the hot-electron temperature $T_{\rm eff}$ for a charge-neutral graphene in the range of
$\sim500$ K, and we
expect the phonon temperature $T_\text{ph}$ to be significantly
lower than $T_{\rm eff}$ and $\omega_\text{ph}$.

Using the above formulation, we directly simulate the nonequilibrium steady-state and compute the steady-state current under given electric fields. After the self-consistent calculation is finished, the current density is calculated with the following definition\cite{jong-prb},
\begin{align}
J=i\gamma\sum_{p_y}\langle d^\dag_{\ell+1,A}(p_y)d^\dag_{\ell,B}(p_y)-\text{H.c.}\rangle/a_y.
\end{align}
We refer the readers to the literature for more details\cite{jiajun-prb,jiajun-prl,jiajun-nano}.

\section{Results}

\subsection{Signature of Landau-Zener tunneling}
\begin{figure}
\includegraphics[scale=0.5]{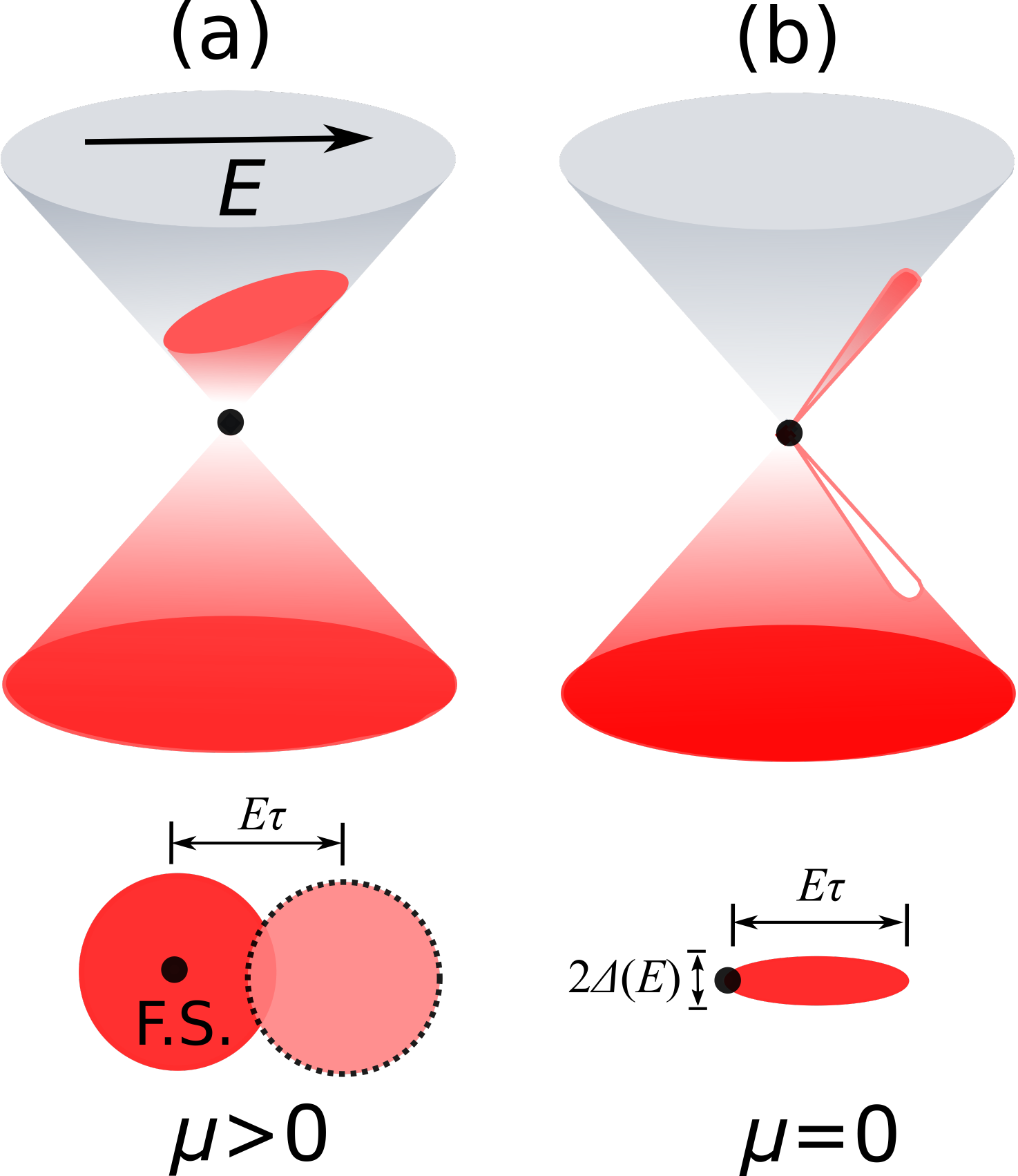}
\caption{Schematic demonstration of momentum distribution of
electric-field-driven graphene (a) off the Dirac point and (b)
on the Dirac point. (a) Chemical potential $\mu>0$ and a finite Fermi
sphere exists in equilibrium. An electric field displaces the Fermi
sphere by $E\tau$ according to the semiclassical picture. (b) $\mu=0$
and the ``Fermi sphere'' becomes pointlike. In this case, any finite
field transport effect is of quantum mechanical nature and should be
related to creation of particle-hole pairs. A
jetlike distribution for both electrons and holes is created under
a finite electric field in the charge-neutral limit.}
\label{sketch}
\end{figure}

We first consider the strong-field transport in graphene without optical
phonon interaction. The transport behavior is quite different between
the cases with zero and finite chemical potential $\mu$. With $\mu>0$, a
non-zero Fermi circle exists in the upper band, allowing the
semiclassical Boltzmann transport theory to be applied: electric fields
displace the Fermi circle in the field direction, as shown in the
Fig.~\ref{sketch}(a). With a scattering-time $\tau$, the Fermi circle is
displaced by $E\tau$. A similar argument can be made for holes in the case
of $\mu<0$. With $\mu=0$ the Fermi circle shrinks to a point, as in
Fig.~\ref{sketch}(b), and the conventional linear-response theory does
not apply. Near this point, an electric field excites electrons to the
upper band, leaving holes in the lower band. These nonequilibrium
excitations are initially driven by the Landau-Zener transition, and the
electrons are accelerated by the electric field during the scattering
time $\tau$, resulting in a highly anisotropic excitation distribution.
Physically, the steady-state current is established when the
Landau-Zener tunneling and field-driven acceleration of electrons are
balanced by electron-phonon interaction as well as other dephasing
mechanisms. The excitation distribution in the momentum space forms
streaking lines in the particle and hole cones along the direction of
the field. This is a solid-state analog of the Schwinger
effect~\cite{schwinger51}, a particle-antiparticle pair creation by an
intense electric field with zero mass gap.

\begin{figure}
\includegraphics[scale=0.55]{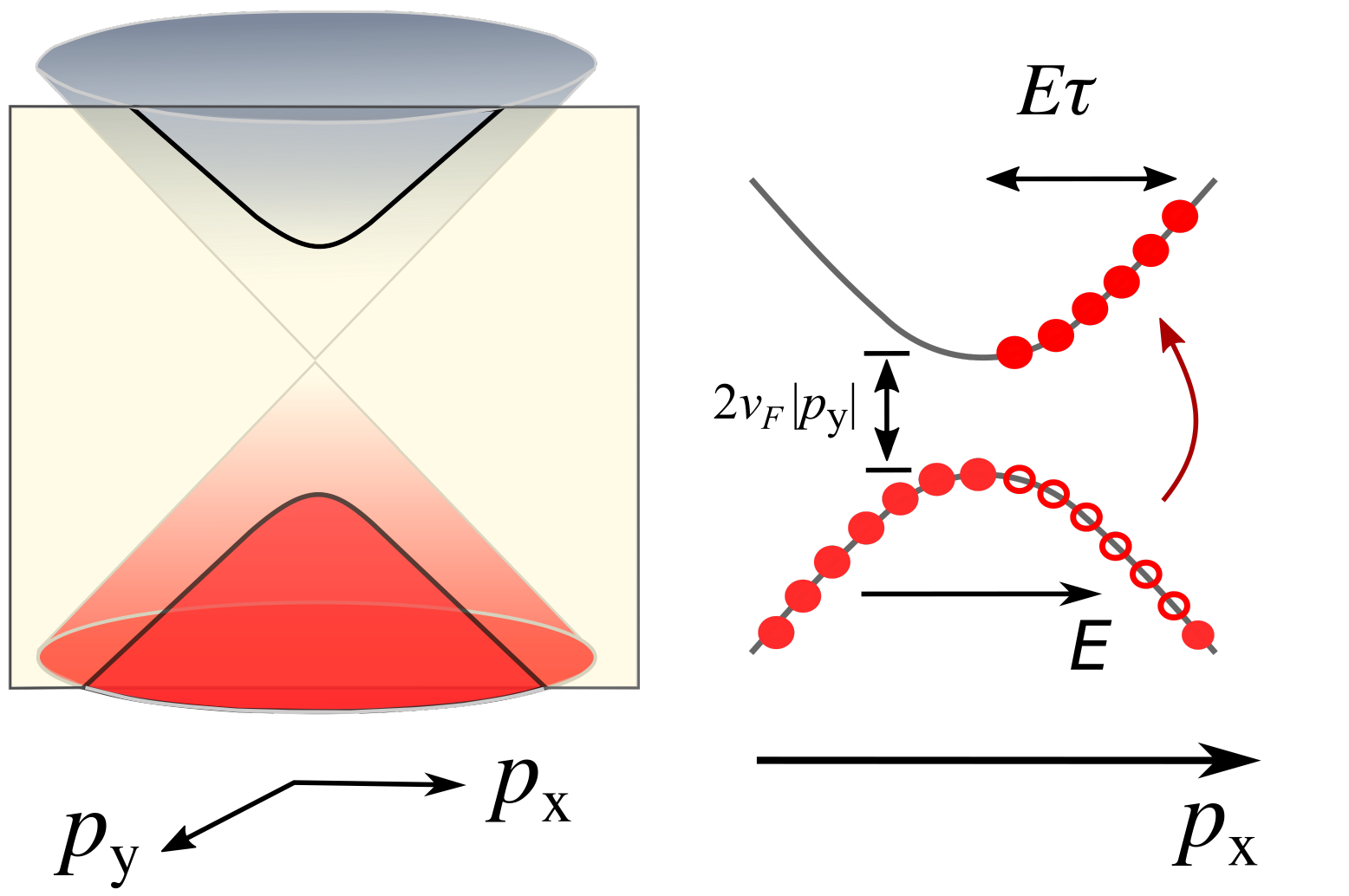}
\caption{Demonstration of Landau-Zener tunneling in graphene on the
Dirac point. When the Dirac cone intersects with a plane of nonzero
$p_y$, the excitations in the gapped quasi-one-dimensional dispersion relation is
understood in terms of the Landau-Zener tunneling. Significant
electron-hole pairs are created for $|p_y|\lesssim\Delta(E)$.
}
\label{LZsketch}
\end{figure}

With an electric field in the $x$ direction, the perpendicular momentum
$p_y$ is a good quantum number and we analyze the excitations on the
dispersion relation on a sliced cone at a fixed $p_y$, as shown in
Fig.~\ref{LZsketch}. At $p_y\neq 0$, the dispersion relation is gapped
with the charge gap $2v_F|p_y|$. The transition of an electron
from the lower band to the upper band due to a constant force field is
described by the Landau-Zener tunneling with the transition probability
$\gamma_{\rm LZ}$ as
\begin{align}
\gamma_\text{LZ}=\exp(-\pi v_Fp_y^2/E).
\end{align} 
This suggests that electrons only with $|p_y|\lesssim \sqrt{E/\pi
v_F}\equiv \Delta_0(E)$ are excited to the upper band [$\Delta_0(E)$ is
the width of the distribution under electric field
$E$] with the
subscript $0$ referring to without phonons. On the other hand,
the range of excitation of the longitudinal momentum $p_x$ is given by
the lifetime of the excited electrons. The fermion bath provides the
lifetime~\cite{jong-prb,mitra08} $\tau_\Gamma=(2\Gamma)^{-1}$ and the range of $p_x$ for excited
electrons become $0\lesssim p_x\lesssim E\tau_\Gamma$. Combining these
observations, an ansatz is proposed for the momentum distribution
for the excitations as
\begin{align}
n_{\bm{p}}\propto\theta(\Delta_0(E)-|p_y|)\theta(E\tau_\Gamma-p_x)\theta(p_x),
\label{nkansatz}
\end{align}
and the number of excited electrons $n_{\rm ex}$ behaves as
\begin{equation}
n_{\rm ex}\propto \Delta_0(E)E\tau_\Gamma \propto E^{3/2}/\Gamma.
\end{equation}
The excitation of holes has exactly the same distribution as the
electrons, with the jet-like excitation on the same side of the
momentum.
Similar momentum distribution has been proposed in the streaming
model~\cite{fang11}, where the Fermi sea is elongated in the direction
of the field by $E\tau$. However, in the $\mu=0$ limit, a finite Fermi
sea does not exist and this phenomenological approach does not provide
any mechanism for the width of the stream.

\begin{figure}
\centering
\includegraphics[scale=0.35]{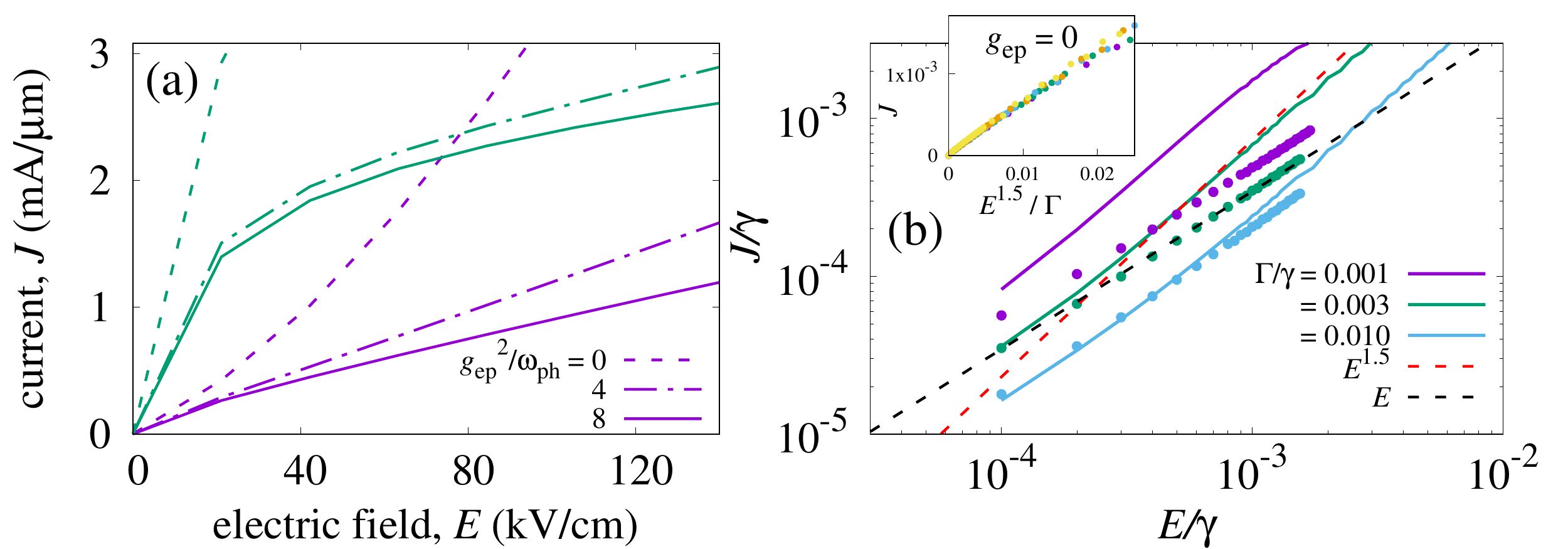}
\caption[$J$-$E$ relation of graphene under strong
field]{$J-E$ relation of graphene under strong field. (a) Deviations of
current from non-interacting case ($g_{\rm ep}=0.0$, dash lines) to
$g_{\rm ep}^2/\omega_\text{ph}=4\gamma$ (dotted dash lines) and
$8\gamma$ (solid lines). The damping $\Gamma=0.001\gamma$. The
green (purple) lines are with $\mu=0.10\gamma$
($\mu=0.0\gamma$). (b) The currents with (dots) and
without (solid lines) the optical-phonon interactions at different
damping parameter $\Gamma$. As damping increases, the range of
electric field in which $J\propto E^{1.5}$ holds expands. The inset
shows the current as a function of $E^{1.5}/\Gamma$ for a variety of $\Gamma
= 0.001,0.003,0.005,0.007,0.100$. In (b), $g_{\rm
ep}^2/\omega_\text{ph}=4\gamma$ for dots. Optical phonon frequency
$\omega_\text{ph}$ is $0.05\gamma$. In all plots in this paper, $\gamma$
is set to $3$ eV. }
\label{jegrph}
\end{figure}

This simple argument leads to a straightforward prediction of the $I$-$V$ characteristics. 
In the limit of $E\tau_\Gamma\gg \Delta_0(E)$, the jet-like distribution
is almost completely aligned with the electric field and the averaged
velocity is close to $v_F$, resulting in $J\propto
n_\text{neq}v_F\propto E^{1.5}\tau_\Gamma$, which is verified in
Fig.~\ref{jegrph}. In the absence of the el-ph coupling
($g_{\rm ep}=0$), the $J\propto E^{1.5}$ scaling law is shown (solid lines)
with the power greater than 1 as the signature of Landau-Zener
mechanism. The inset shows the collapse of data to the form $J\sim E^{1.5}/\Gamma$.

When the optical phonon interaction is turned on ($g_{\rm ep}\neq 0$), the
superlinear $JE$ relation becomes marginally sublinear.
The strongly inelastic scattering by optical phonons reduces the current,
as shown in Fig.~\ref{jegrph}(a). It is interesting that the window for
the $J\sim E^{3/2}$ shrinks as the damping $\Gamma$ is reduced as shown in Fig.~\ref{jegrph}(b), whereas
it may be naively expected that a clean limit ($\Gamma\to 0$)  may
preserve the peculiar fractional power law. This can be explained as
follows. In the small $\Gamma$ limit, the electrons lifetime increases
and the energy increase due to the acceleration $v_F(E\tau_\Gamma)$
reaches the optical phonon threshold $\hbar\omega_{\rm ph}$ at a smaller
field $E$. This is consistent with the findings that the superlinear
behavior is observed in low-mobility devices under dc-electric
field~\cite{Vandecasteele10}, as well as graphene samples excited by THz
electric pulses~\cite{oladyshkin17}. In the clean limit, the observed $JE$
relations remain close to linear in strong-field transport
experiments~\cite{barreiro09,Yang18}. 

\subsection{Steady-state current at large fields}

\begin{figure}
\includegraphics[scale=0.35]{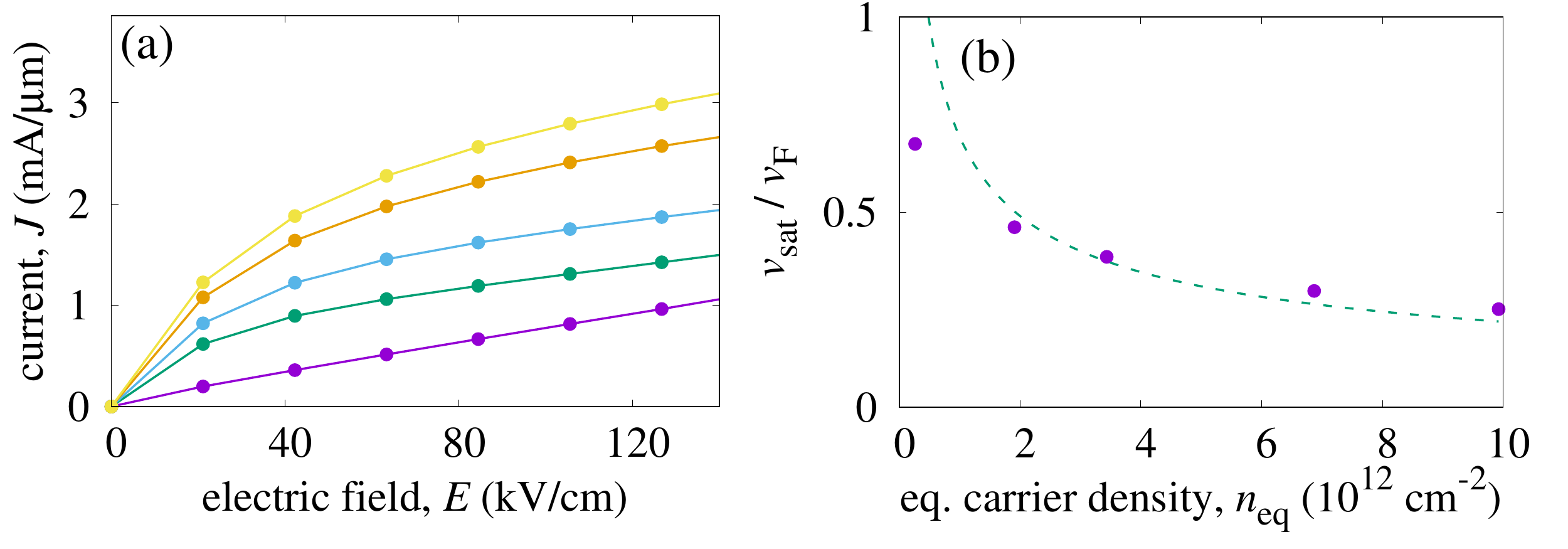}
\caption{(a) $J$-$E$ curve for different chemical potential $\mu$. As
system is taken away from the Dirac point,  current increases and shows
the saturating behavior under high-field. The $J$-$E$ curves in (a) have chemical potentials $\mu=0.0,0.05,0.07,0.10,0.12$ from bottom to top. (b) Computed
relation (dots) $v_{\rm sat}\propto 1/\sqrt{n_{\rm eq}}$ between the saturated velocity $v_\text{sat}$ and the
equilibrium charge density $n_\text{eq}$. This is
consistent with the semiclassical formula \eqref{empirical} (dashed
line).
$\Gamma=0.003\gamma$, $\omega_\text{ph}=0.05\gamma$, and
$g_\text{ep}^2/\omega_\text{ph}=8\gamma$. }
\label{current}
\end{figure}

Now we study the effect of optical phonon interaction more closely.
Recently, the phenomena of current saturation have attracted intense
interest.\cite{shishir09, dorgan10,dasilva10,perebeinos10,ramamoorthy15} It
is found that the current saturation in graphene at high electric field depends on
the optical phonon frequency $\omega_\text{ph}$ and equilibrium current
carrier density $n_\text{eq}$. The former is a parameter independent of
electric control,
and the latter is controlled through the chemical potential $\mu$. We are mostly
interested in how the graphene behaves under different chemical
potentials, especially when it is close to the Dirac point $\mu=0$.
Figure~\ref{current} systematically shows calculated $J\sim E$ relations with a
set of realistic parameters. In graphene, the tight-binding
hopping parameter $\gamma\approx 3$ eV and the optical-phonon
frequency $\omega_\text{ph}=0.05\gamma\approx 150$ meV. These values are
close to the empirical parameter used in previous works~\cite{fang11,ramamoorthy15,perebeinos10}.

In Fig.~\ref{current}, the electronic current generally shows the
tendency to saturate under high electric
fields in the samples with relatively high equilibrium electron density.
Previous semi-classical analyses~\cite{meric08,fang11} have assumed that
Fermi sphere is shifted by $\hbar\omega_\text{ph}/v_F$, leading to an
empirical formula of the saturated velocity, 
\begin{align}
v_\text{sat}=\frac{1}{\sqrt{\pi}}\frac{\omega_\text{ph}}{\sqrt{n_\text{eq}}},
\label{empirical}
\end{align}
where $n_\text{eq}$ is the equilibrium current carrier density. While
this expression has been confirmed experimentally, the
formula obviously breaks down when $\omega_\text{ph}$ is too large or
$n_\text{eq}$ is too small, since $v_\text{sat}$ can never be greater
than $v_F$. Our main interest is in the regime the approximation fails.
Our calculations do not show true saturation of current and we derive
the drift velocity according to a phenomenological model~\cite{perebeinos10}:
\begin{align}
v_d=\frac{\chi_0 E}{1+\chi_0 E/v_\text{sat}},
\end{align}
where $\chi_0$ is the zero-field mobility. As verified in Fig.~\ref{current}(b), the extracted $v_\text{sat}$
follows $1/\sqrt{n_\text{eq}}$ relation until the Dirac point is
reached. Interestingly, recent measurements using short bias
pulses~\cite{ramamoorthy15} reported a similar range of the maximum drift
velocity $v_d\approx 0.5\times v_F$ close to the Dirac point.

\subsection{Evolution of momentum distribution under external field}
\label{evol}

\begin{figure}
\centering
\includegraphics[scale=0.3]{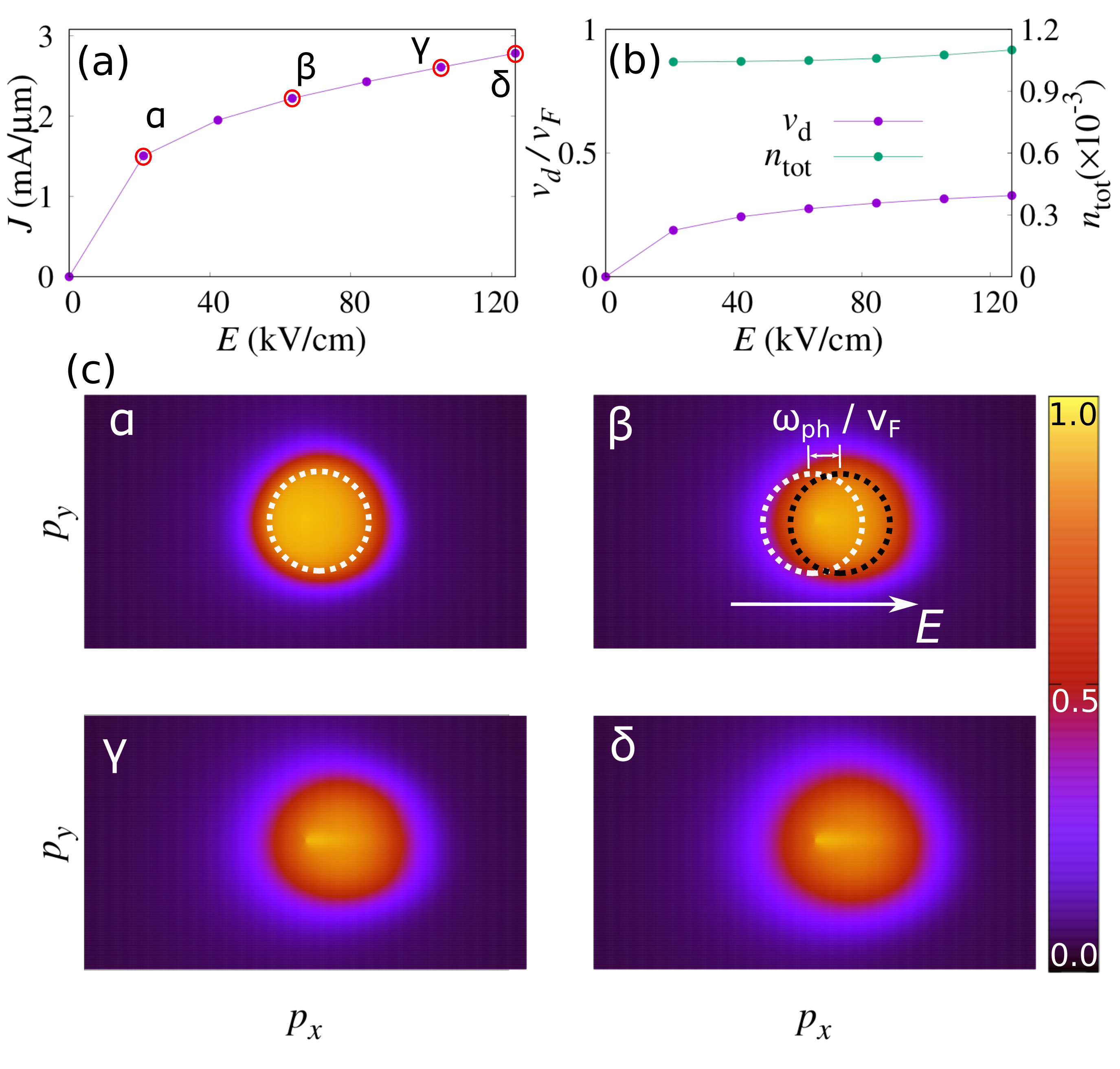}
\caption{Charge excitations and the momentum distribution of
electrons under electric fields in the off-Dirac-point graphene. (a) Saturation of current under electric fields and (b) drift velocity $v_d$
and total current carriers number $n_\text{tot}$ under electric fields.
The $v_d$ saturates like current, and $n_\text{tot}$ is almost
unchanged. (c) Momentum distributions of electrons at $\mu=0.1\gamma$. 
Fermi sea is shifted at small electric fields. Its displacement is nearly
unchanged at high fields. Near ${\bf p}=0$, a faint signature of the
Schwinger effect can be seen in the panels $\gamma$ and $\Delta$. }
\label{nk10}
\end{figure}

To further understand the steady-state current due to the optical phonon
scattering, we look at the evolution of momentum distribution
$n_{\bm{p}}$ under electric fields. The formulation is detailed in Appendix A.
At $\mu>0$ a finite Fermi sea
exists around the center of the Dirac cone, as shown in Fig.~\ref{nk10}
for the current and momentum distributions. The Fermi sea is shifted
along the field-direction when electric field is applied. However, at
high electric fields, the Fermi sea stops to shift due to the strong
relaxation by the optical phonons when electrons lose their excess energy to
phonon baths. This is consistent with the semi-classical Boltzmann
transport calculations.\cite{chauhan09}

\begin{figure}
\centering
\includegraphics[scale=0.3]{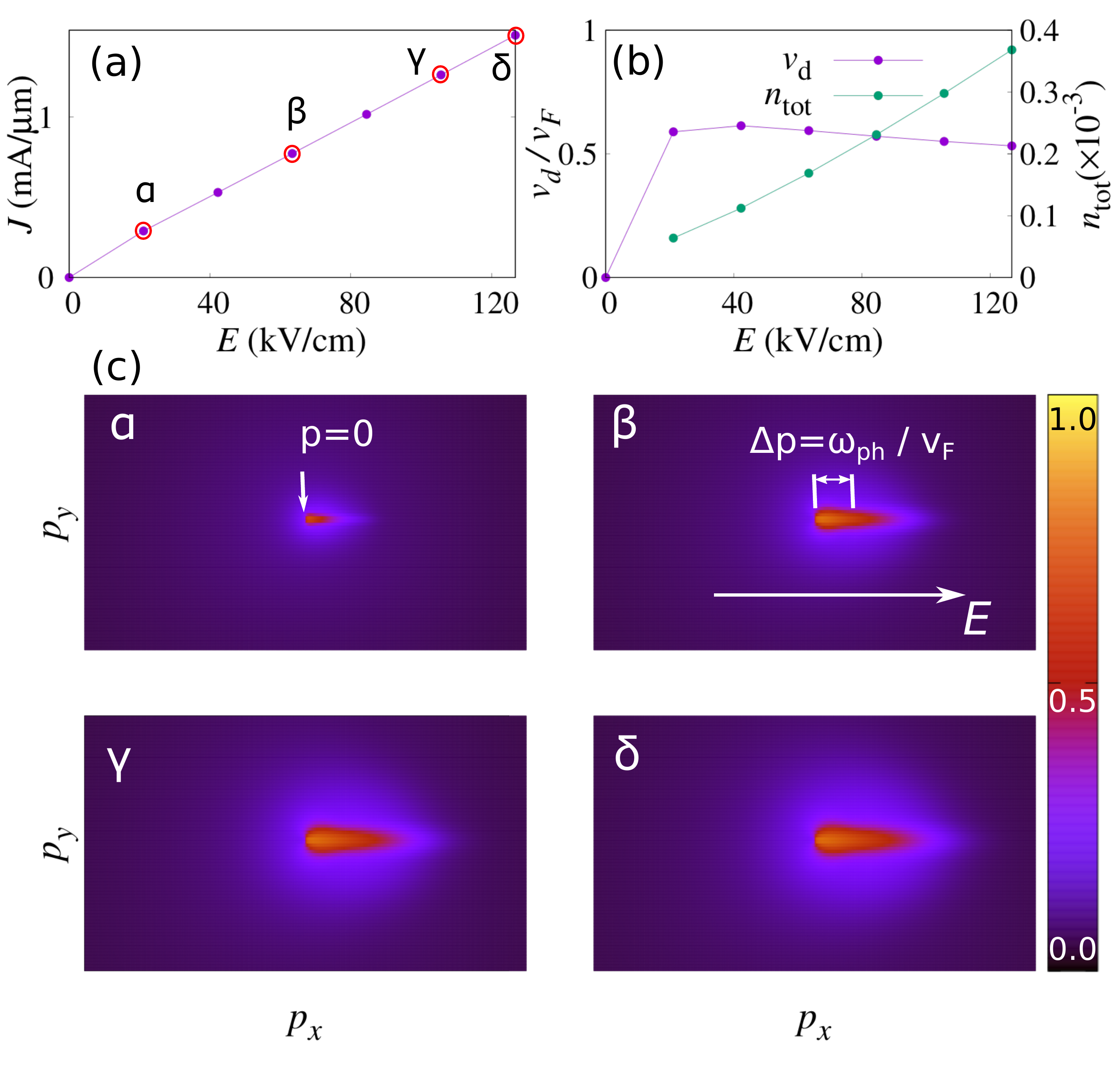}
\caption[Current and momentum distribution at Dirac
point]{Charge excitations and the momentum distribution of electrons
under electric fields in the on-Dirac-point graphene.  (a) Current which scales linearly
with electric field. (b) Drift velocity $v_d$ and total number of
current carriers $n_\text{tot}$. The drift velocity overshoots and then
decreases due to optical phonon emission, and the total number of
excitations $n_\text{tot}$ increases monotonically with $E$.  
(c) Momentum distribution of electrons (upper band). At
high fields, the center of excitation distribution is at the momentum $p_x=\Delta
p\approx \omega_{\rm ph}/v_F$. The distribution of holes is essentially identical. }
\label{nk00}
\end{figure}

At $\mu=0$, as shown in Fig.~\ref{nk00}(a), the current increases almost
linearly without saturation. To investigate the origin, we plot the
electron excitation $n_{\rm ex}$ and the drift velocity $v_d$ calculated
by
\begin{align}
v_d=v_F\frac{\int_{|{\bm p}|<\Lambda_p}n({\bm p})\!\cos\theta
d^2{\bm p}}{\int_{|{\bm p}|<\Lambda_p}n({\bm p})
d^2{\bm p}},
\label{vdintegral}
\end{align}
with the angle $\theta$ between the field and momentum vectors, and the
large momentum cutoff $\Lambda_p$. It is clear that the drift velocity
saturates at small field and the main contribution to the current
$J=n_{\rm ex}v_d$ is due to the electron excitations linearly
proportional to the field $E$. We emphasize that the drift velocity is
evaluated by dividing the current by the excited charge out of
neutrality, instead of the equilibrium charge $n_{\rm eq}$ as often used
in experimental estimates.

Let us investigate more into the excitations in the momentum space.
Figure~\ref{nk00}(c) shares the characteristics proposed in
Eq.~(\ref{nkansatz}). However, coupling to the inelastic phonons results
in important differences. First, as the electric field increases, the
length of the jet stops growing but saturates to the length related
to the phonon frequency, $(p_x)_{\rm max}\sim
2\omega_{\rm ph}/v_F$  with the center of distribution
at $\omega_{\rm ph}/v_F$.
Therefore, $E\tau_\Gamma$ in Eq.~(\ref{nkansatz}) is replaced by
$2\omega_{\rm ph}/v_F$. Second, the width of the distribution $\Delta(E)$
no longer follows the Landau-Zener form $\Delta_0(E)\propto\sqrt{E}$. As
the distribution saturates, the width $\Delta(E)$ gets smeared due to
the local scattering by optical phonons, which can be summarized as
\begin{align}
n_{\bm{p}}\propto\theta(\Delta(E)-|p_y|)\theta(2\omega_{\rm ph}/v_F-p_x)\theta(p_x).
\label{nk}
\end{align}
Due to the smearing of the momentum distribution with stronger
el-ph scattering at high fields, the drift velocity is
slightly reduced with a wider angular distribution in Eq.~(\ref{vdintegral}), 
as shown in Fig.~\ref{nk00}(b).

We now discuss the role of the electron-electron scattering in the $I$-$V$
characteristics considered in this section\cite{fang11,li10}. We expect
that the qualitative nature of the charge excitations discussed above
remain robust against the $e$-$e$ interactions. Due to the momentum and
energy conservation in the Coulomb scattering\cite{brida13} in graphene,
the phase space in the scattering process is strongly limited. Out of a
jetlike distribution [see Fig.  \ref{nk00}(c)], intra-band
scattering events with collinear incoming momenta will scatter into
another pair of collinear momenta moving in the same direction,
preserving the strong anisotropy of the distribution. Other interband
e-e interaction processes, such as carrier multiplication and the Auger
process, should also be suppressed\cite{brida13} in this case since they
involve scattering electrons (holes) into fully occupied (empty) states.
This argument does not apply in ultrafast
measurements~\cite{malic17,higuchi17}, in which electrons incoming with
opposite momenta induced by oscillating field can scatter into any
outgoing direction and relax to an isotropic momentum distribution.

\subsection{Energy dissipation}

\begin{figure}
\centering
\includegraphics[scale=0.35]{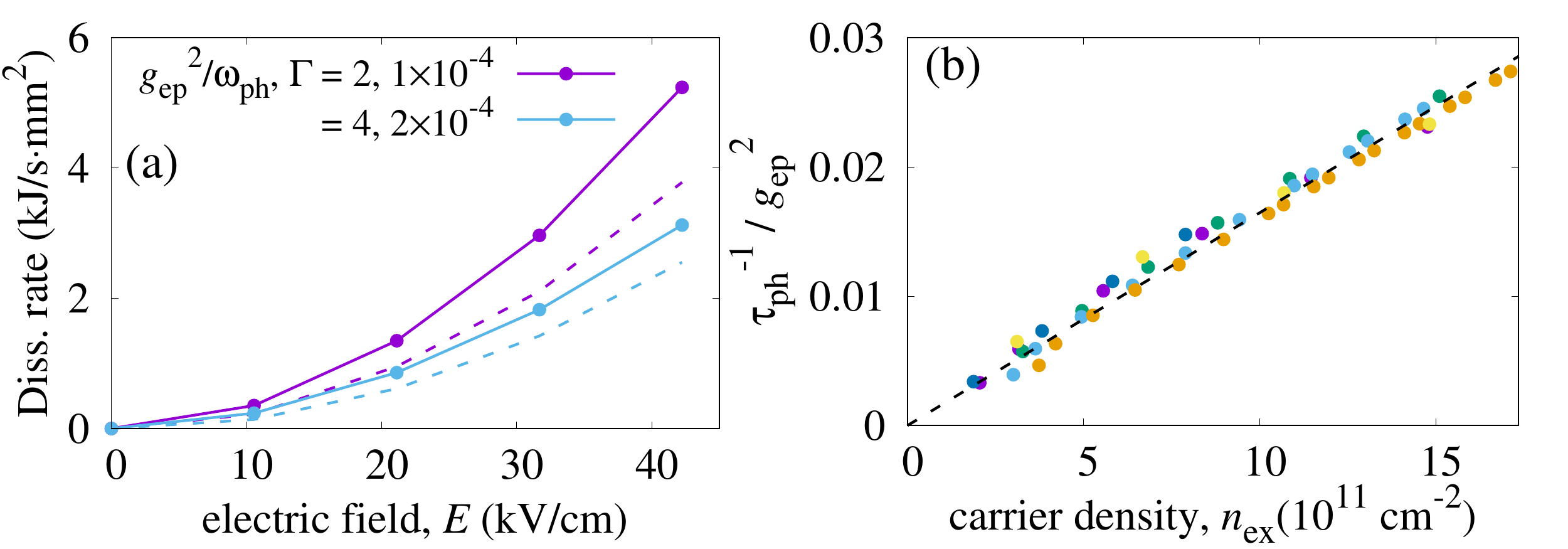}
\caption{(a) Energy dissipation rate in the presence of optical
phonons with frequency $\omega_\text{ph}$, and (b) the observed
$\tau^{-1}_\text{ph}\propto n_\text{ex}$ relation. In (a), the dots are
total dissipation rate $JE$, and the dashed lines are the dissipation
rates due to optical phonon emission, predicted by Eq.~\eqref{disseq}.
The predicted dissipation rate does not deviate much from the numerically calculated values at strong optical phonon interaction. We have assumed
$\hbar\omega_\text{ph}\approx 150$ meV and $\Gamma\sim
0.0001\gamma$, which corresponds to
$\tau^{-1}_\Gamma\sim 1\text{ ps}^{-1}$ assuming $\gamma=3$ eV. The
$g_{\rm ep}^2/\omega_\text{ph}\sim 2\gamma$ leads to $\tau^{-1}_\text{ph}$
being in the order of $10\text{ ps}^{-1}$. In (b), the linear relation of $\tau^{-1}=\alpha n_\text{ex}$ is tested for a variety of interaction strengths $g^2_\text{ep}/\omega_\text{ph}=2,4,8,12,16$ and $\Gamma=0.0001,0.0002,0.001$. }
\label{diss}
\end{figure}

To understand the strong-field $IV$ relation at Dirac point with
optical-phonon interaction, we look into the energy conservation law:
the electric power $JE$ is equal to the dissipation rate. In the case
when the optical phonon scattering dominates at large $E$, the
scattering rate is
$\tau^{-1}_\text{ph}=-\text{Im}(\Sigma^<_\text{ph}-\Sigma^>_\text{ph})$
and each scattering between an electron and optical phonon reduces the
electron energy by $\hbar\omega_\text{ph}$. So the dissipation rate of
nonequilibrium excitations is
$n_\text{ex}\omega_{\text{ph}}\tau^{-1}_\text{ph}$. The system at finite
temperature can have a small equilibrium current carrier density
$n_\text{eq}$, but we will concentrate on the $\mu=0$ case where 
$n_\text{eq}\approx0$, and the total electron density $n_\text{tot}\approx 
n_\text{ex}$. We then have
\begin{align}
JE&=n_\text{ex}\omega_{\text{ph}}\tau^{-1}_\text{ph}+\Theta_\Gamma.
\label{disseq}
\end{align}
The $\Theta_\Gamma$ is the dissipation rate due to fermion reservoirs.
In the following, we will focus on the case where $\Theta_\Gamma$ is
physically negligible. This approximation is tested in Fig.~\ref{diss}(a), with $\tau_\text{ph}$ being the scattering rate at
$\omega=0$.

The el-ph scattering rate
$\tau^{-1}_{\rm ph}$ is given as a convolution of electron and phonon
Green's functions and is thus proportional to the electron
density. We show that
\begin{equation}
\tau_\text{ph}^{-1}=\alpha n_\text{ex},
\label{tau2alpha}
\end{equation}
as in Fig.~\ref{diss}(b),
with the coefficient $\alpha$ explicitly calculated in
Appendix B. It is shown both theoretically and numerically that $\alpha$ is proportional to $g_\text{ep}^2$, given by $\alpha = \pi A_c g_\text{ep}^2/2\omega_\text{ph}$. The parameter $\alpha$ is independent of
the electric field, and it can be evaluated experimentally at zero
field by the line broadening of the electrons due to phonons.

%\section{The relation of $\tau^{-1}=\alpha n_\text{tot}$}

A crucial step to understand the linear $J$-$E$ relation is the relation
$n_{\rm ex}\propto E$. We may relate the el-ph scattering rate
$\tau_{\rm ph}^{-1}$ in two different ways. One can argue that, in the
high-field limit, the frequency of the el-ph scattering is determined by
how fast the electron energy gained by the $E$-field reaches the optical
phonon frequency, that is, with the average energy excitation at momentum $\Delta p$ [see
Fig.~\ref{nk00}(c)] given as $v_F\Delta p=v_F E\tau_{\rm
ph}\sim\omega_{\rm ph}$,
which leads to $\tau_{\rm ph}^{-1}\sim v_FE/\omega_{\rm ph}$.
This result is consistent with a more careful analysis based on Eq.~\eqref{disseq}. In the saturated drift velocity limit, the current $J$
in the energy-conservation relation \eqref{disseq} is replaced by
$n_\text{ex}v_\text{sat}$. We then re-derive
\begin{equation}
\tau^{-1}_\text{ph}=v_\text{sat}E/\omega_\text{ph}.
\label{tau2vsat}
\end{equation}
after eliminating $n_\text{ex}$. 

\begin{figure}
\centering
\includegraphics[scale=0.5]{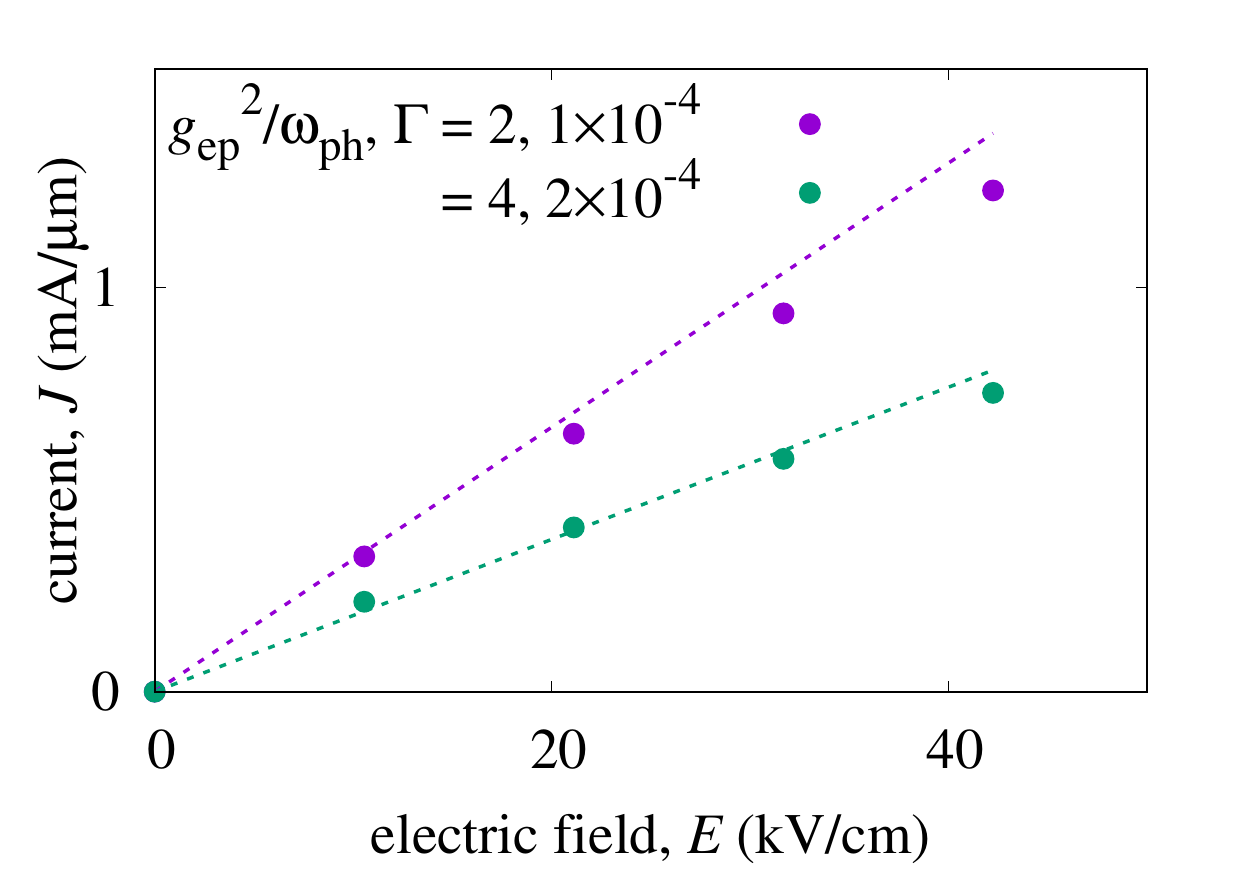}
\caption{$J$-$E$ relation of Dirac electron compared with the theoretical
prediction. The theoretical $J$-$E$ relation \eqref{curr} is compared
against the numerical data (dots). $v_\text{sat}\sim0.5v_F$ in the
formula is extracted from the numerical data. The theory correctly
predicts that the power of $J$-$E$ relation is close to one, and
quantitatively matches the numerical data. Bath temperature is set to
35K in this calculation.}
\label{currfor}
\end{figure}

By eliminating $\tau_{\rm ph}^{-1}$ from the Eqs.~\eqref{tau2alpha} and
\eqref{tau2vsat}, we obtain
\begin{align}
n_\text{ex}= \frac{v_\text{sat}}{\alpha\omega_\text{ph}}E,
\end{align}
and the current-field relation immediately follows
close to the Dirac point (after restoring the physical constants)
\begin{align}
J\approx n_\text{ex}
v_\text{sat}=\frac{e^2v_\text{sat}^2}{{\alpha\hbar\omega_\text{ph}}}E.
\label{curr}
\end{align}
The saturated velocity has weak dependence on the electric field in the
strong field limit. This result shows that the main electric field
dependence originates from the charge excitation proportional to $E$.
The formula is tested with numerical results in Fig.~\ref{currfor}. The
parameter $v_\text{sat}$ is extracted from numerical calculations. In
experiments, the formula may provide a way to extract $v_\text{sat}$
from the measured $I$-$E$ relation in the presence of both optical phonon
emission and Landau-Zener tunneling.  Using the relation $n_{\rm
ex}\tau_{\rm ph}=\alpha^{-1}$, the $J$-$E$ relation can be cast in the
usual Drude form as $J=(n_\text{ex}\tau_\text{ph}/m^*)E$ with the
effective mass $m^{*-1}\sim v_\text{sat}^2/\hbar\omega_\text{ph}$ for
the driven Dirac particles, whose kinetic energies are of the magnitude
of $\hbar\omega_\text{ph}$.

We reemphasize that, despite its similarity to the Ohmic law of simple metals, the origin of
the above linear $J$-$E$ relation is very different. In a simple metal, while the
carrier density $n_\text{eq}$ is weakly perturbed by the electric
fields, the drift velocity $v_d$ is proportional to the $E$-field. For
the Dirac electrons in graphene, however, the role of the electric field is
reversed: $v_d\approx v_\text{sat}\sim v_F$ at saturation while the
nonequilibrium carriers density $n_\text{ex}\propto E$. 

\subsection{Effective temperature}
A nonequilibrium effective temperature in a dc-transport system is the
direct consequence of the balance between the electric power and the
energy dissipation. Here, we follow the procedure in
Ref.~\citenum{jiajun-nano} to define the effective temperature from the nonequilibrium distribution function
\begin{align}
f_\text{loc}(\omega)=-\frac{\text{Im}G_\text{loc}^<(\omega)}{2\text{Im}G_\text{loc}^r(\omega)}.
\end{align}
The nonequilibrium distribution function usually has a different
functional form from the equilibrium Fermi-Dirac distribution in which
temperature is a well-defined parameter. To obtain a comparable
nonequilibrium temperature parameter, we define the effective
$T_\text{eff}$ from the first moment of the distribution function,
\begin{align}
\frac{\pi^2}{6}T^2_\text{eff}=\int \omega\left[f_\text{loc}(\omega)-\theta(-\omega+\mu)\right]
d\omega.
\end{align}
$\Theta(\omega)$ is the Heaviside step function. This definition is consistent with the Fermi-Dirac distribution at an equilibrium temperature.

\begin{figure}
\centering
\includegraphics[scale=0.5]{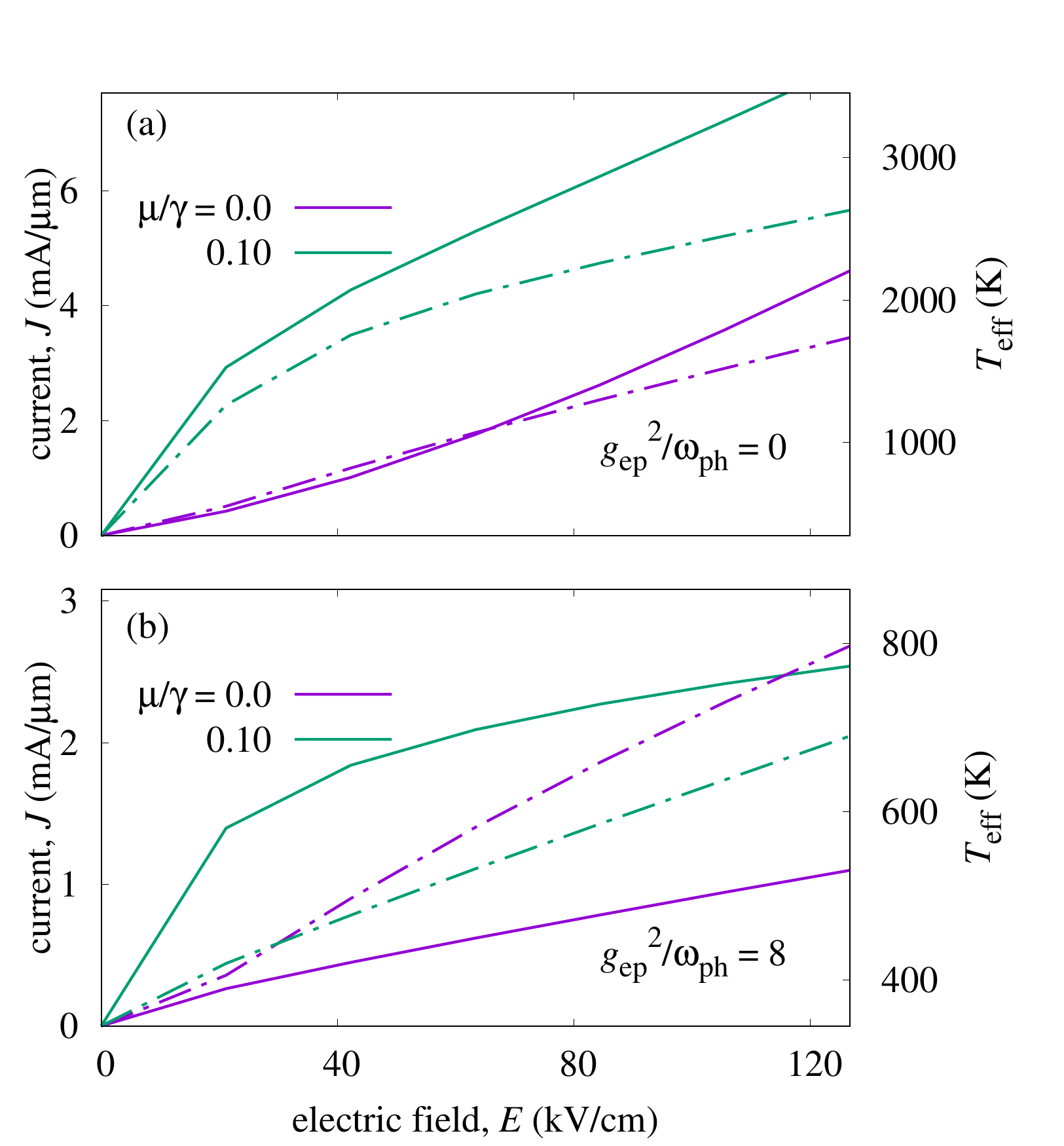}
\caption{Current and effective temperature of graphene (a) on and (b) off Dirac point. $\Gamma=0.001\gamma$. The solid lines are currents, while dot-dashed lines are \emph{nonequilibrium} effective temperature calculated under different electric fields. When optical-phonon coupling is switched off, the sample with $\mu=0.10t$ (away from the Dirac point) shows higher effective temperature due to higher current as well as Joule heating; however, optical phonon emission results in current saturation as discussed above, and the behavior of $T_\text{eff}$ becomes more complicated, with a crossover of $T_\text{eff}$ occurring at finite electric fields. }
\label{teff}
\end{figure}

With this definition, we plot the effective temperature in
Fig.~\ref{teff}. In the case of $g_{\rm ep}=0$, effective temperature of
the off-Dirac-point graphene is always higher than that of the on-Dirac-point system. This is because higher current carrier density results in higher current. The system with more current-carrying excitations create more Joule heating thereby the temperature is also higher. However, this is dramatically changed when optical phonon interaction is considered. In this case, the system with higher electron density still has higher electric current for all electric fields. However, at higher electric fields, the effective temperature of the off-Dirac-point system falls below the temperature of the Dirac-electron system. 

To explain this crossover behavior, we note that the effective
temperature is the measure of the amount of excitations. As the electric
field increases, the el-ph coupling overrides the relaxation by the
fermion baths. As we argued above, the energy relaxation by phonons
is proportional to the electronic charge density, and at non-zero $\mu$, the finite
Fermi surface makes the relaxation more efficient, thereby cools the
electron temperature more effectively than at the $\mu=0$ limit. The
effective temperature is predicted to be in the range of 400-800 K in the
realistic range of the electric
field.\cite{freitag10,bae10,bae11,luxmoore13,beechem16} We note that
recent works~\cite{ferry17,Yang18} have reported much higher
hot-electron temperatures in graphene systems.

\section{Conclusion}

We have demonstrated that Dirac electrons in graphene become excited in
a non-trivial manner and lead to a marginally linear dependence of
electric current under a dc electric field, through microscopic
calculations based on the Keldysh Green's-function theory. Inelastic
scattering by optical phonons is shown to be crucial to produce the
nonequilibrium charge excitation population $n_{\rm ex}$ and the
electric current linearly
proportional to the external field, close to a Dirac point,
as summarized as
\begin{equation}
J\approx \frac{e^2v_\text{sat}^2}{\alpha \hbar\omega_\text{ph} }E,
\end{equation}
with the saturated drift velocity $v_{\rm sat}$, the optical phonon
frequency $\omega_{\rm ph}$, and the coefficient $\alpha$ proportional to the
el-ph coupling constant.
The mechanism for this apparent Ohmic $I$-$V$ characteristics in the Dirac
limit is different from the conventional Drude model in that the linear
dependence of the electric field comes from the nonequilibrium charge
density instead of the drift velocity.
This linear $JE$ relation
without saturation close to the Dirac point has been observed in clean
graphene encapsulated by the
hexa-boron-nitrides~\cite{Yang18,yamoah}.
Away from the Dirac
charge-neutrality point, the conventional Boltzmann transport is recovered
with the tendency for the current saturation with where the
drift velocity is proportional to $\omega_{\rm
ph}/n_{\rm eq}^{1/2}$ with the equilibrium charge density $n_{\rm eq}$.

The electron-hole excitations at the charge-neutral point ($\mu=0$) are
strongly anisotropic in the momentum space with the excited electrons
and holes staying in the momentum regime of the same direction, an analog of the Schwinger
effect in a solid state system. The inter-band creation of electron-hole
pairs poses possibilities for optical applications. The
continuous excitation energy up to the optical phonon energy of $\sim 150$
meV can be used for infra-red optical devices, without any lower
threshold. Most intriguing is the change of
the coupling of the electron-hole pairs to photons
upon switching of the bias. Close to the Dirac point, the wavefunction
under dc electric field acquires extra phase oscillation due to the
potential gradient which can strongly suppress the photon-generation.
When the electric-field is turned off, the electron wavefunction becomes a
plane-wave and the coupling to the photon field enhances roughly by the
factor proportional to the length of the sample.
This mechanism may be utilized in a fast-switching IR-optic diode.

\section{Acknowledgements}
JL acknowledges the computational support from the Center for Computational Research (CCR) at University at
Buffalo. We thank Jonathan P. Bird, Takuya Higuchi, Alexander Khaetskii, Huamin Li and Christian Heide for their helpful discussions
throughout the study.

\appendix
\section{Momentum distribution of electrons}
To compute the momentum distribution of electrons, we note that $\mathbf{n}_{\bm{p}}=-i\mathbf{G}^<_{\bm{p}}(t,t)
=-i\sum_{\bm{r}}\exp(i\bm{p}\cdot\bm{r})\mathbf{G}^<_{\bm{r}\bm{0}}(t,t)\nonumber$. We have used matrix-valued Green's functions and $\mathbf{n}_{\bm{p},ss'}=-iG^<_{\bm{p},ss'}(t,t)$. Using the time-translational invariance of the Green's functions, time $t$ can be fixed as $0$, so the momentum distribution is calculated as
\begin{align}
\mathbf{n}_{\bm{p}}&=-i\sum_{\bm{r}}\exp(i\bm{p}\cdot\bm{r})\int \frac{d\omega}{2\pi} \mathbf{G}^<_{\bm{r}\bm{0}}(\omega)\nonumber\\
&=-i\sum_{\bm{r}}\exp(i\bm{p}\cdot\bm{r})\int \frac{d\omega}{2\pi} \sum_{\bm{r}'}\mathbf{G}^r_{\bm{r}\bm{r}'}(\omega)\mathbf{\Sigma}^<(\omega+\bm{r}'\cdot\bm{E})\nonumber\\
&\quad\times\mathbf{G}^a_{\bm{r}'\bm{0}}(\omega),
\end{align}
where $\mathbf{\Sigma}^<(\omega)=\mathbf{\Sigma}_\Gamma^<(\omega)+\mathbf{\Sigma}_\text{ph}^<(\omega)$ is the total lesser self energy, including components from both fermion reservoirs and optical phonon baths. Now we shift $\omega\to\omega-\bm{r}'\cdot\bm{E}$, and notice that $\mathbf{G}^r_{\bm{r+a}\bm{r'+a}}(\omega)=\mathbf{G}^r_{\bm{r}\bm{r}'}(\omega+\bm{a}\cdot\bm{E})$, with $\bm{r},\bm{r}'$ and $\bm{a}$ being lattice vectors. Therefore the formula is reduced to
\begin{align}
\mathbf{n}_{\bm{p}}&=-\frac{i}{2\pi}\int d\omega\sum_{\bm{r}\bm{r}'}\exp(i\bm{p}\cdot\bm{r})\mathbf{G}^r_{\bm{r}\bm{r}'}(\omega-\bm{r}'\cdot\bm{E})\mathbf{\Sigma}^<(\omega)\nonumber\\
&\quad\times\mathbf{G}^a_{\bm{r}'\bm{0}}(\omega-\bm{r}'\cdot\bm{E})\nonumber\\
&=-\frac{i}{2\pi}\int d\omega\sum_{\bm{r}\bm{r}'}\exp(i\bm{p}\cdot(\bm{r}-\bm{r}'))\mathbf{G}^r_{\bm{r}-\bm{r}',\bm{0}}(\omega)\mathbf{\Sigma}^<(\omega)\nonumber\\
&\quad\times[\exp(-i\bm{p}\cdot\bm{r}')\mathbf{G}^r_{-\bm{r}'\bm{0}}(\omega)]^\dag\nonumber\\
&=-\frac{i}{2\pi}\int d\omega\mathbf{G}^r_{\bm{p}}(\omega)\mathbf{\Sigma}^<(\omega)\mathbf{G}^a_{\bm{p}}(\omega),
\label{nkgrph}
\end{align}
where we have defined $\mathbf{G}^r_{\bm{p}}(\omega)=\sum_{\bm{r}}\exp(i\bm{p}\cdot\bm{r})\mathbf{G}^r_{\bm{r}\bm{0}}(\omega)$. In practical calculations, we firstly compute $\mathbf{G}^r_{\bm{r}\bm{0}}(\omega)$ and Fourier transform them to $\mathbf{G}^r_{\bm{p}}(\omega)$ in momentum space. Then $n_{\bm{p}}$ is calculated by evaluating the integral in \eqref{nkgrph}. Finally, to interpret the result $\mathbf{n}_{\bm{p}}$, we should expand it in terms of equilibrium diagonalized basis\cite{neto09}, 
\begin{align}
\psi_{\pm,\bm{p}}&=\frac{1}{\sqrt{2}}\begin{pmatrix}e^{-i\theta_{\bm{p}}/2}\\\pm e^{i\theta_{\bm{p}}/2}\end{pmatrix},\nonumber\\
&\text{with}\quad\exp(i\theta_{\bm{p}})=\epsilon_{\bm{p}}/|\epsilon_{\bm{p}}|,
\end{align}
with $\epsilon_{\bm{p}}$ given by
\begin{align}
\epsilon_{\bm{p}}=-\gamma\left(1+2 e^{-ip_x a_x}\cos\frac{p_y a_y}{2}\right).
\end{align}
We define unitary transformation $U_{\bm{p}}=\begin{pmatrix}\psi_{+,\bm{p}}&\psi_{-,\bm{p}}\end{pmatrix}$, and transform the $\mathbf{n}_{\bm{p}}$,
\begin{align}
\mathbf{\tilde{n}}_{\bm{p}}=U_{\bm{p}}^\dag\mathbf{n}_{\bm{p}}U_{\bm{p}}
\end{align}
Then the particle numbers for upper/lower bands are $\mathbf{\tilde{n}}_{\bm{p},++}$ and $\mathbf{\tilde{n}}_{\bm{p},--}$. These equations will be useful to calculate the nonequilibrium momentum distribution.

\section{Derivation of Eq.~(\ref{tau2alpha}), $\tau^{-1}_\text{ph}=\alpha n_\text{ex}$}
\label{taun}

In our self-consistent calculations, the 2nd-order electron self-energy
by the optical phonon interaction is given by
\begin{align}
\Sigma^\lessgtr_\text{ph}(\omega)&\approx g_{\rm ep}^2G^\gtrless(\omega\mp\omega_\text{ph}),
\end{align}
with coupling constant $g_{\rm ep}$. The approximation is made due to the nearly empty optical phonon bath, where phonons are rarely excited before the nonequilibrium excitations having energy close to the optical phonon energy. Then the scattering rate at $\omega=0$ becomes
\begin{align}
\tau_\text{ph}^{-1}&=-ig_{\rm ep}^2[G^<(\omega_\text{ph})-G^>(-\omega_\text{ph})]\nonumber\\
&=2\pi g_{\rm ep}^2 [A(\omega_\text{ph})f_\text{loc}(\omega_\text{ph})+A(-\omega_\text{ph})\left(1-f(-\omega_\text{ph})\right)]\nonumber\\
&=4\pi g_{\rm ep}^2\rho(\omega_\text{ph}),
\label{tau-rho}
\end{align}
with the particle-hole symmetry in the spectral function
$A(\omega)=A(-\omega)$ and the distribution function
$f_{\rm loc}(\omega)=1-f_{\rm loc}(-\omega)$ at the charge-neutrality point $\mu=0$.
$\rho(\omega)=A(\omega)f_\text{loc}(\omega)$ is the occupation density
of electrons at frequency $\omega$. As a first-order approximation, we again
assume the jet-like distribution $n_{\bm{p}}$ is uniform within a thin
rectangular box aligned in the field-direction and zero outside the box,
as in Eq.~\eqref{nk}. 
Note the number of excited electrons in the energy interval $[\omega,\omega+d\omega]$ should be 
\begin{align}
\rho(\omega)d\omega\propto \int_{\omega<v_Fp<\omega+d\omega}
n_{\bm{p}}d^2\bm{p}.
\end{align}
This rectangle-shaped momentum distribution results in the
uniform $\rho(\omega)\approx\rho(\omega_\text{ph})$ when $|\omega|\lesssim 2\omega_\text{ph}$ and zero otherwise. Then we have
\begin{align}
n_\text{ex}=2\int{d\omega\rho(\omega)}/A_c\approx 4g_v \rho(\omega_\text{ph})\omega_\text{ph}/A_c,
\label{nex-rho}
\end{align}
with $g_v=2$ counting the valley degrees of freedom and
$A_c=\frac{3\sqrt{3}}{2}a^2$ being the area of unit cell. The prefactor
$2$ is included to count both electrons and holes. By comparing
\eqref{tau-rho} and \eqref{nex-rho}, we have 
\begin{align}
\tau_\text{ph}^{-1}&=\alpha n_\text{ex},\quad\text{with
}\alpha=\frac{\pi g_{\rm ep}^2 A_c}{2\omega_\text{ph}}.
\label{alphap}
\end{align}

\bibliography{grph-ref.bib}

\end{document}